\documentclass[conference]{IEEEtran}
\IEEEoverridecommandlockouts
\usepackage{amsmath}
\usepackage{amssymb}
\usepackage{amsfonts}
\usepackage{amsthm}
\usepackage{algorithm}
\usepackage{algorithmic}
\usepackage{graphicx}
\usepackage{array}

\usepackage[shortlabels]{enumitem}
\usepackage{cite}
\usepackage{color}
\usepackage{epsfig,epstopdf}
\usepackage{url}
\usepackage{subfigure}
\usepackage{balance}
\usepackage{multirow}
\usepackage[letterpaper,left=0.68in,right=0.68in,top=0.7in,bottom=1.05in]{geometry}

\IEEEoverridecommandlockouts

\DeclareMathOperator*{\argmin}{arg\,min}

\begin{document}

\title{Few-Shot Transfer Learning for Device-Free Fingerprinting Indoor Localization}

\author{\IEEEauthorblockN{Bing-Jia Chen and Ronald Y. Chang}
\IEEEauthorblockA{Research Center for Information Technology Innovation, Academia Sinica, Taiwan}
\IEEEauthorblockA{Email: b07901088@ntu.edu.tw, rchang@citi.sinica.edu.tw}
\thanks{This work was supported in part by the Ministry of Science and Technology, Taiwan, under Grant MOST 109-2221-E-001-013-MY3.}}

\maketitle

\begin{abstract}
Device-free wireless indoor localization is an essential technology for the Internet of Things (IoT), and fingerprint-based methods are widely used. A common challenge to fingerprint-based methods is data collection and labeling. This paper proposes a few-shot transfer learning system that uses only a small amount of labeled data from the current environment and reuses a large amount of existing labeled data previously collected in other environments, thereby significantly reducing the data collection and labeling cost for localization in each new environment. The core method lies in graph neural network (GNN) based few-shot transfer learning and its modifications. Experimental results conducted on real-world environments show that the proposed system achieves comparable performance to a convolutional neural network (CNN) model, with 40 times fewer labeled data.
\end{abstract}


\section{Introduction}

Indoor localization plays an important role in many Internet of Things (IoT) applications. A wireless approach to indoor localization is attractive since it is economical and nonintrusive. Fingerprint-based techniques using wireless signals are widely adopted \cite{academia}, and can be classified as either device-based or device-free depending on whether a tracking device is required to be attached to the target. Deep learning has found success in fingerprint-based techniques to analyze the signal measurements. Deep learning-based fingerprinting systems have been proposed for device-based \cite{RSSI,CSI-based,CNN} and device-free \cite{DNN} indoor localization using the received signal strength indicator (RSSI) and/or channel state information (CSI) of the wireless signals.

A key challenge to deep learning-based fingerprinting approaches for wireless indoor localization is data labeling. This includes the need to collect timely labeled data in one environment as the environment dynamics may change over time (e.g., for real-time localization), and the need to collect a new set of labeled data for localization in each new environment (e.g., for multi-environment localization). This time-sensitive and environment-dependent nature of wireless data incurs significant data collection and maintenance cost for supervised learning which requires a large amount of labeled data. To address this problem, semi-supervised learning and transfer learning have been considered. Semi-supervised learning relies on a small amount of labeled data and a large amount of unlabeled data for model training, thus reducing the data labeling cost. A generative adversarial network (GAN) based semi-supervised learning scheme was proposed \cite{DCGAN}, where the model can be trained with unlabeled data as well as artificial data generated by GAN. Variational auto-encoder (VAE) based schemes were also investigated \cite{VAE,Chen21}.

Transfer learning adopts a slightly different approach to reducing the data labeling cost. The idea is to reuse a large amount of labeled data previously collected in other environments ({\it source domains}) and newly collect only a small amount of labeled data in the current environment ({\it target domain}). This approach however faces several challenges, such as redundant knowledge from the source domains, limited amount of data in the target domain, and environment heterogeneity \cite{TranLoc}. A heterogeneous knowledge transfer framework was proposed to improve the robustness of fingerprint-based localization against the environmental dynamics for device-based systems \cite{TranLoc}. A transfer learning framework was proposed for fingerprint-based localization to reduce the offline training overhead by reshaping data distributions in the target domain based on the transferred knowledge from the source domains \cite{low-overhead}. In these works, the main objective was to address the environmental dynamics. Besides, device-based systems using RSSI were considered and some prior knowledge of the target domain was typically required. Since device-free systems using CSI are known to be much more sensitive to environmental changes as compared to device-based systems using RSSI, conventional domain knowledge transferring techniques may not work effectively and new transfer learning techniques may be needed for the new scenario.

In this work, we introduce the concept of few-shot learning \cite{GNN_few_shot} into transfer learning for device-free indoor localization using CSI. Traditionally, few-shot learning aims at training a general meta-model that can adapt to all kinds of tasks quickly using only few data for each new task \cite{matching, prototypical, meta-learning}. In our application, the ``tasks'' are more specific, which are localization in various heterogeneous environments. By combining few-shot learning and transfer learning, we propose to add both a large number of labeled CSI samples from the source domain and a small number of labeled CSI samples from the target domain to the training dataset. The support and query sets in each learning task are formed by sampling the training dataset. Training is performed by analyzing the relations among samples in the support and query sets and minimizing the classification errors for the samples in the query set. A graph neural network (GNN) based approach is adopted to analyze the relations among samples due to its superior performance compared with conventional few-shot learning models \cite{GNN_few_shot}. The main contributions of this paper are:
\begin{itemize}
    \item We propose a general method to reduce the data collection/labeling cost in fingerprint-based indoor localization by reusing the labeled data previously collected in other environments and transferring the model learned therein to a new environment. The proposed method is applicable to any settings of the source and target domains, including layouts, dimensions, and numbers of locations.
    \item The proposed method is based on GNN few-shot transfer learning and its enhancements. The proposed method demonstrates remarkable performance with as few as $1$, $5$, or $10$ labeled CSI samples per location in the target domain, which is more than $40$ times reduction of labeled data as compared to the convolutional neural network (CNN) model trained with $400$ labeled CSI samples per location in the target domain.
\end{itemize}

The outline of the paper is as follows. Sec.~\ref{sec:II} describes the problem, challenges, and motivations. Sec.~\ref{sec:GNN} presents the proposed GNN-based few-shot transfer learning scheme and its modifications. Sec.~\ref{sec:result} presents the performance results and discussion. Finally, Sec.~\ref{sec:conclusion} concludes the paper.

\section{Problem Description} \label{sec:II}

We consider the device-free fingerprint-based indoor localization problem and model it as a classification problem. The objective is to determine the unknown location of a target person out of $N$ possible locations in some indoor environment. As in a typical fingerprint-based approach, a site survey is performed in the offline training phase to build the fingerprint database (signal map) and the measured signals in the online testing phase are matched with the fingerprint database to determine the location of the target in our approach. A tracking device is not required to be attached to the target, i.e., device-free. Our fingerprints are channel state information (CSI) measurements. We use the Intel Wi-Fi Wireless Link 5300 802.11n multiple-input multiple-output (MIMO) radios to collect CSI samples \cite{CSItool}. Each CSI sample is a $W=30\times 2\times 2=120$ dimensional vector, representing $30$ orthogonal frequency division multiplexing (OFDM) subcarriers over a $2\times 2$ MIMO channel.

There are several challenges in the fingerprint-based approach. First, collecting a large set of labeled data to build the fingerprint database is time-consuming and labor-intensive. Second, the fingerprint database could be dated due to changes and dynamics in the environment over time, and thus it may need to be updated from time to time. Third, the fingerprint database is environment-dependent, and a new and independent site survey is required for each new environment. To address these challenges, we develop a localization scheme that can significantly reduce the data labeling effort and fingerprint maintenance/construction costs for localization in multiple, different indoor environments. The proposed scheme exploits the concept of few-shot learning \cite{few-shot} and transfer learning \cite{transfer}. More specifically, we aim to reuse the existing data in one environment with $M$ locations ({\it source domain}) to aid the localization in another environment with $N$ (where in general $N\neq M$) locations ({\it target domain}), with only few newly collected labeled data in the target domain. The objective is to achieve localization in the target domain as accurately as possible. 

\section{GNN-Based Few-Shot Transfer Learning for Device-Free Indoor Localization} \label{sec:GNN}

In this section, we elaborate on the proposed scheme whose core method lies in GNN-based few-shot transfer learning and its modifications.

\begin{figure}[t]
\centerline{\includegraphics[width=\columnwidth]{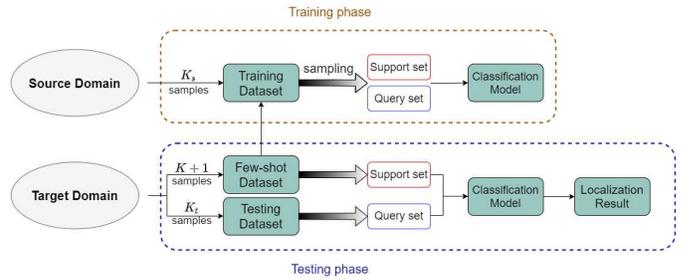}}
\caption{The schematic of few-shot transfer learning based indoor localization.}
\label{system}
\end{figure}

\subsection{System Overview}

\begin{figure*}[t]
\begin{center}
\includegraphics[width=0.85\textwidth]{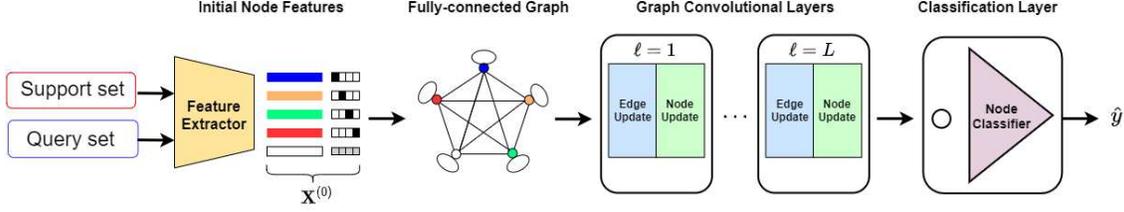}
\end{center}
\caption{GNN-based few-shot learning.}
\label{few-shot}
\end{figure*}

The general framework of our proposed scheme is an $N$-way $K$-shot few-shot learning model incorporating the concept of transfer learning. The schematic of the proposed few-shot transfer learning for indoor localization is shown in Fig.~\ref{system}. In the training phase, the training dataset consists of existing data from the source domain ($K_s$ samples per location) and few newly collected data from the target domain ($K+1$ samples per location, where $K+1\ll K_s$). Thus, the training dataset contains a total of $K_s\times M+(K+1)\times N$ samples. In each learning task $\mathcal{T}$, there is an episode comprised of a {\it support set} and a {\it query set} which share the same label space. The support set comprises $N$ classes with $K$ samples for each class. The $N$ classes are randomly selected from the total $M+N$ classes (locations) in both source and target domains combined. The query set comprises one sample which is to be classified as one of the classes in the support set. The classification of the sample in the query set is supervised and the result is used to optimize the model. The model used is a GNN-based few-shot learning model which will be described in the next subsection. The reason we need $K+1$ samples per location from the target domain is to accommodate both the support and query sets. 

In the testing phase, the $N$ classes in the support set are all from the $N$ classes in the target domain, with $K$ samples for each class which are randomly selected from the {\it same} $K+1$ samples used in the training. Thus, no additional data samples are needed prior to online testing in the testing phase. The testing samples are fed, one by one, into the query set for online testing. A testing dataset of $K_t$ samples per location ($K_t\times N$ samples in total) from the target domain is collected to evaluate the performance of the model.

\subsection{GNN-Based Few-Shot Learning} \label{sec:originalGNN}

In this and following subsections, we describe the classification model in Fig.~\ref{system} which is realized by GNN-based few-shot learning or its variant. The schematic of GNN-based few-shot learning is shown in Fig.~\ref{few-shot}. The four main components are described as follows.

\subsubsection{Feature Extractor}

For a learning task $\mathcal{T}$ in $N$-way $K$-shot learning, let $K_{\rm total}$ be the total number of CSI samples in the support set and query set, where $K_{\rm total}=KN+1$ for both training and testing phases. Let ${\mathbf x}_i, i=1,\ldots, K_{\rm total}$ be a CSI sample in the support or query set. Each ${\mathbf x}_i$ is associated with a $N\times 1$ label ${\mathbf y}_i=[y_1,y_2,\ldots,y_N]^T$. The label ${\mathbf y}_i$ is a one-hot encoded vector with a single $1$-element which corresponds to the true class of this sample and all-zero elements elsewhere if ${\mathbf x}_i$ is in the support set, and ${\mathbf y}_i=[1/N,\dots,1/N]^T$ which represents a uniform distribution over the label space if ${\mathbf x}_i$ is in the query set. A feature extractor $\phi:\mathbb{R}^{W\times 1} \mapsto \mathbb{R}^{d\times 1}$ is applied on ${\mathbf x}_i$. The initial features of ${\mathbf x}_i$ are given by
\begin{equation} \label{initial node features}
{\mathbf x}_i^{(0)} = [\phi({\mathbf x}_i)^T, \, {\mathbf y}_i^T]^T
\end{equation}
where $\mathbf x_i^{(0)} \in \mathbb{R}^{d_0\times 1}$ with $d_0=d+N$. The initial features of all ${\mathbf x}_i$'s are collectively expressed as ${\mathbf X}^{(0)}=\big[{\mathbf x}_1^{(0)}, \ldots, {\mathbf x}_{K_{\rm total}}^{(0)}\big]^T \in {\mathbb R}^{K_{\rm total}\times d_0}$.

\subsubsection{Fully-Connected Graph}

A fully-connected weighted undirected graph ${\cal G}=({\cal V}, {\cal E})$ is constructed, where each vertex/node $v_i\in {\cal V}$ ($i\in \{1,2,\ldots,K_{\rm total}\}$) represents a CSI sample ${\mathbf x}_i$ and each edge (including self-loop) $e_{ij}\in {\cal E}$ ($i,j \in \{1,2,\ldots,K_{\rm total}\}$) is associated with a weight $w_{ij}$. Initially, the node features for the $i$th node are ${\mathbf x}_i^{(0)}$.

\subsubsection{Graph Convolutional Layers}

The edge weights and node features are updated and the similarities between any two nodes are learned through graph convolutional layers. The number of graph convolutional layers, denoted by $L$, is a design parameter. Each graph convolutional layer $\ell$ ($\ell=1,2,\ldots, L$) comprises two sequential steps, i.e., {\it edge update} and {\it node update}. For the edge update, first define an adjacency matrix $\mathbf{A}^{(\ell)} \in \mathbb{R}^{K_{\rm total} \times K_{\rm total}}$ on ${\cal G}$, whose $(i,j)$-th element is denoted by $A_{ij}^{(\ell)}=\big(\mathbf{A}^{(\ell)}\big)_{ij}$. The edge update in the $\ell$th ($\ell=1,2,\ldots, L$) convolutional layer calculates
\begin{equation} \label{edge update}
A_{ij}^{(\ell)} = f_{\theta}\left(\left\vert \mathbf x_i^{(\ell-1)} - \mathbf x_j^{(\ell-1)}\right\vert\right)
\end{equation}
where $\mathbf{x}_i^{(\ell-1)}\in \mathbb{R}^{d_{\ell-1} \times 1}$ denotes the node features of the $i$th node at the input of the $\ell$th convolutional layer, $|\cdot |$ denotes the absolute value, and $f_{\theta}: \mathbb{R}^{d_{\ell-1}\times 1} \mapsto \mathbb{R}$ is an edge update function with $\theta$ being learnable parameters. At last, softmax is applied on each row of updated adjacent matrix $\mathbf{A}^{(\ell)}$ to make sure that $A^{(\ell)}_{ij}\in [0,1]$. The node update in the $\ell$th ($\ell=1,2,\ldots, L$) convolutional layer first calculates
\begin{equation} \label{eq:node update1}
\mathbf{X}'^{(\ell-1)} = \rho(\mathbf{A}^{(\ell)}\mathbf{X}^{(\ell-1)}\mathbf{W}^{(\ell-1)})
\end{equation}
where $\mathbf{X}^{(\ell-1)}=\big[{\mathbf x}_1^{(\ell-1)}, \ldots, {\mathbf x}_{K_{\rm total}}^{(\ell-1)}\big]^T\in \mathbb{R}^{K_{\rm total} \times d_{\ell-1}}$ is the node feature matrix at the input of the $\ell$th convolutional layer, $\mathbf{W}^{(\ell-1)}\in \mathbb{R}^{d_{\ell-1} \times d'_{\ell-1}}$ is an edge transformation matrix which is trainable, and $\rho$ is the activation function, e.g., ReLU or LeakyReLU. Then, to retain earlier memory, $\mathbf{X}^{(\ell-1)}$ is appended to $\mathbf{X}'^{(\ell-1)}$ to complete the node update, i.e.,
\begin{equation} \label{eq:node update2}
\mathbf{X}^{(\ell)} = \big[\mathbf{X}'^{(\ell-1)}, \mathbf{X}^{(\ell-1)}\big]
\end{equation}
where $\mathbf{X}^{(\ell)}\in \mathbb{R}^{K_{\rm total}\times d_{\ell}}$, with $d_{\ell}=d'_{\ell-1}+d_{\ell-1}$.

\subsubsection{Classification Layer}

The sample in the query set is classified based on the final node feature matrix $\mathbf{X}^{(L)}$. Specifically, \eqref{edge update} is rerun for $\ell=L+1$ where $\mathbf{X}^{(L)}$ is taken as the input of \eqref{edge update} to generate the adjacency matrix $\mathbf{A}^{(L+1)}$. Without loss of generality, let the node corresponding to the sample in the query set be the $K_{\rm total}$-th node on ${\cal G}$. Then, the final predicted result $\widehat{\mathbf{y}}$ for the sample in the query set is obtained by running \eqref{eq:node update1} for $\ell=L+1$ but with the activation function $\rho$ replaced by a row-wise softmax function denoted by $\sigma$, and extracting the $K_{\rm total}$-th row. This can be expressed as
\begin{equation} \label{classification layer}
\widehat{\mathbf{y}} = \sigma\Big(\mathbf{W}^{(L)}\big(\mathbf{A}^{(L+1)}\mathbf{X}^{(L)}\big)_{K_{\rm total}}^T\Big) \in \mathbb{R}^{N\times1}
\end{equation}
where ${\mathbf W}^{(L)}\in \mathbb{R}^{N\times d_{L}}$ is a trainable matrix to map the node features to the distribution in the label space. Let $\widehat{y}_n$ be the probability that the sample in the query set is in the $n$th class and $\widehat{\mathbf{y}}=[\widehat{y}_1,\widehat{y}_2,\ldots,\widehat{y}_N]^T$. The cross-entropy loss used in the training phase is given by
\begin{equation} \label{cross entropy}
\mathcal{L}_{\rm GNN} = -\sum_{n=1}^{N} y_n \log \widehat{y}_n.
\end{equation}

GNN-based few-shot learning could have the over-smoothing problem \cite{over-smoothing}. This is mainly because the node update is based on the information from all nodes in the fully-connected graph structure. Thus, to further enhance the classification performance, we propose to apply three methods to constrain the information exchanged among neighbor nodes, resulting in three modifications to the original GNN-based few-shot learning, as described in Secs.~\ref{sec:AttentiveGNN}--\ref{sec:ChebyNet}, respectively.

\subsection{Attentive GNN} \label{sec:AttentiveGNN}

In GNN-based few-shot learning, the weights in the adjacency matrix represent the degrees of similarity between any two nodes. The idea of attentive GNN \cite{Attentive_GNN} is to sparsify the adjacency matrix to keep only the most important elements. This way, the information exchanged among neighbor nodes is constrained and the over-smoothing problem can be mitigated. A sparse adjacency matrix $\widehat{\mathbf A}^{(\ell)}$ is obtained by solving
\begin{equation} \label{attentive}
\widehat{\mathbf A}^{(\ell)} = \argmin_{{\mathbf B}} \left\| {\mathbf B} - {\mathbf A}^{(\ell)}\right\| \quad \mbox{s.t. } \left\| {\mathbf b}_i \right\|_0 \leq \beta K_{\rm total}
\end{equation}
where ${\mathbf b}_{i}$ is the $i$th row of ${\mathbf B}$, $\left\|\cdot \right\|_0$ is the $\ell_0$ pseudo-norm which counts the number of nonzero elements of its argument, and $\beta\in(0,1]$ is an adjustable parameter that controls the ratio of preserved edge weights. Since \eqref{attentive} is computationally challenging in practice, for simplicity, we keep the largest $\beta K_{\rm total}$ elements in each row of ${\mathbf A}^{(\ell)}$ and set other elements to zero to form the sparse adjacency matrix $\widehat{\mathbf A}^{(\ell)}$ in the training phase. The resulted $\widehat{\mathbf A}^{(\ell)}$ will replace ${\mathbf A}^{(\ell)}$ to be the edge update result and the input for the node update.

\subsection{Edge-Labeling GNN (EGNN)} \label{sec:EGNN}

Edge-labeling \cite{EGNN,Fuzzy_GNN} is employed to combat the over-smoothing problem. The idea is to introduce a regularization term in the loss function of GNN that captures the binary relations among any two different nodes such that the edges connecting nodes from the same class are intensified and the edges connecting nodes from different classes are diminished. Specifically, for the graph ${\cal G}$, define the (true) edge label for the edge connecting the $i$th node and the $j$th node as
\begin{equation} \label{edge label}
\alpha_{ij} = 
\begin{cases}
1, & \text{ if } {\mathbf y}_i = {\mathbf y}_j \\
0, & \text{ if } {\mathbf y}_i \neq {\mathbf y}_j
\end{cases},
\quad i,j\in \{1,2,\ldots,K_{\rm total}\}.
\end{equation}
That is, two nodes that belong to the same class will have an edge label of one, and zero otherwise. Besides, self-loops will have an edge label of one (i.e., $\alpha_{ii}=1, \forall i$). Take $\mathbf{A}^{(L+1)}$ and apply the row-wise softmax function $\sigma$, i.e., $\sigma\big(\mathbf{A}^{(L+1)}\big)$, and let the edge weight $w_{ij}=\big(\sigma\big(\mathbf{A}^{(L+1)}\big)\big)_{ij}$. Note that $w_{ij}$ can be viewed as the probability of predicting that the $i$th node and the $j$th node belong to the same class. Thus, we can define the binary cross-entropy loss
\begin{equation}
\mathcal{L}_{\rm E} = - \frac{1}{\vert \cal{E} \vert}\sum_{i=1}^{K_{\rm total}}\sum_{j=i}^{K_{\rm total}} \alpha_{ij} \log w_{ij} + (1-\alpha_{ij}) \log (1-w_{ij})
\label{binary cross entropy}
\end{equation}
where $|\cdot|$ denotes the cardinality of a set. The overall loss function for EGNN is defined as
\begin{equation} \label{total loss}
\mathcal{L}_{\rm EGNN} = \mathcal{L}_{\rm GNN} + \gamma \mathcal{L}_{\rm E}
\end{equation}
where $\gamma$ is an adjustable parameter. 

\subsection{ChebyNet} \label{sec:ChebyNet}

The idea here is to replace the original node update step at the $\ell$th convolutional layer by the ChebyNet \cite{ChebyNet} and update the node features in the graph spectral domain. Define the normalized weighted graph Laplacian $\mathbf{L}\in \mathbb{R}^{K_{\rm total} \times K_{\rm total}}$ on ${\cal G}$, whose $(i,j)$-th element is
\begin{equation} \label{laplacian}
L_{ij} = 
\begin{cases}
1, & \text{ if } i=j \\
\frac{-w_{ij}}{\sqrt{w_i w_j}}, & \text{ if } i\neq j \text{ and } w_i w_j \neq 0 \\
0, & \text{ otherwise }
\end{cases}
\end{equation}
where $w_{ij}=A_{ij}^{(\ell)}$ with $w_{ii}=0$, and $w_i=\sum_j w_{ij}$ and $w_j=\sum_i w_{ji}$ are the sum of the $i$th row and $j$th row of ${\mathbf A}^{(\ell)}$, respectively. To reduce the learning complexity, we apply the concept of fast localized spectral filtering by the recursive property of Chebyshev polynomial \cite{ChebyNet}. First, define the modified graph Laplacian $\widetilde{\mathbf{L}}= 2\mathbf{L}/\lambda_{\max} - \mathbf I$ to meet the requirement that the domain of the Chebyshev polynomial is in $[-1,1]$. Also, define the input sequence of node features $\widetilde{\mathbf{X}}=\big[\widetilde{\mathbf{X}}_1,\ldots,\widetilde{\mathbf{X}}_n]\in \mathbb{R}^{K_{\rm total}\times(n d_{\ell})}$, where $\widetilde{\mathbf{X}}_1=\mathbf{X}^{(\ell-1)}$, $\widetilde{\mathbf{X}}_2=\widetilde{\mathbf{L}}\mathbf{X}^{(\ell-1)}$, and $\widetilde{\mathbf{X}}_n = 2\widetilde{\mathbf{L}}\widetilde{\mathbf{X}}_{n-1} - \widetilde{\mathbf{X}}_{n-2}$ for $n\geq 3$. Note that $n$ is a parameter that can be chosen from $\{1,2,3,\ldots\}$, which constrains the information exchanged between nodes with $n$-localized property. Then, the node update in the $\ell$th ($\ell=1,2,\ldots, L$) convolutional layer in the graph spectral domain performs
\begin{equation} \label{chebynet}
\mathbf{X}^{(\ell)} = \big[\rho(\widetilde{\mathbf{X}}\widetilde{\mathbf{W}}^{(\ell-1)}),\mathbf{X}^{(\ell-1)}\big]
\end{equation}
where $\widetilde{\mathbf{W}}^{(\ell-1)}\in\mathbb{R}^{(n d_{\ell-1})\times d'_{\ell-1}}$ is a trainable spectral filter, and $\rho$ is the activation function.

\section{Results and Discussion} \label{sec:result}

\subsection{Experimental Settings}

\begin{figure}[t]
\begin{center}
\subfigure[]{
    \label{exp1 FP}
    \includegraphics[width=0.42\columnwidth]{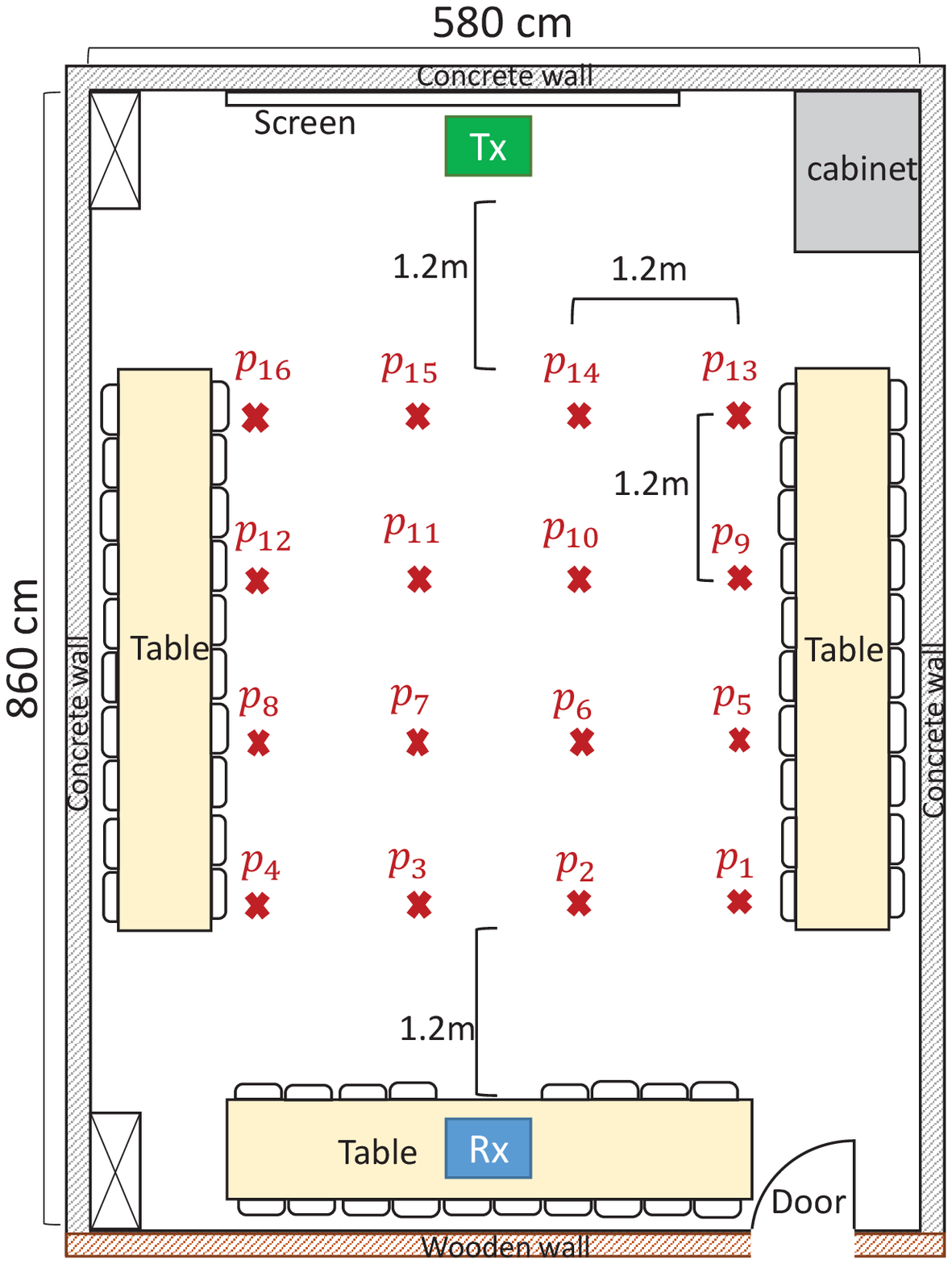}}
\subfigure[]{
    \label{exp1 photo}
    \includegraphics[width=0.42\columnwidth]{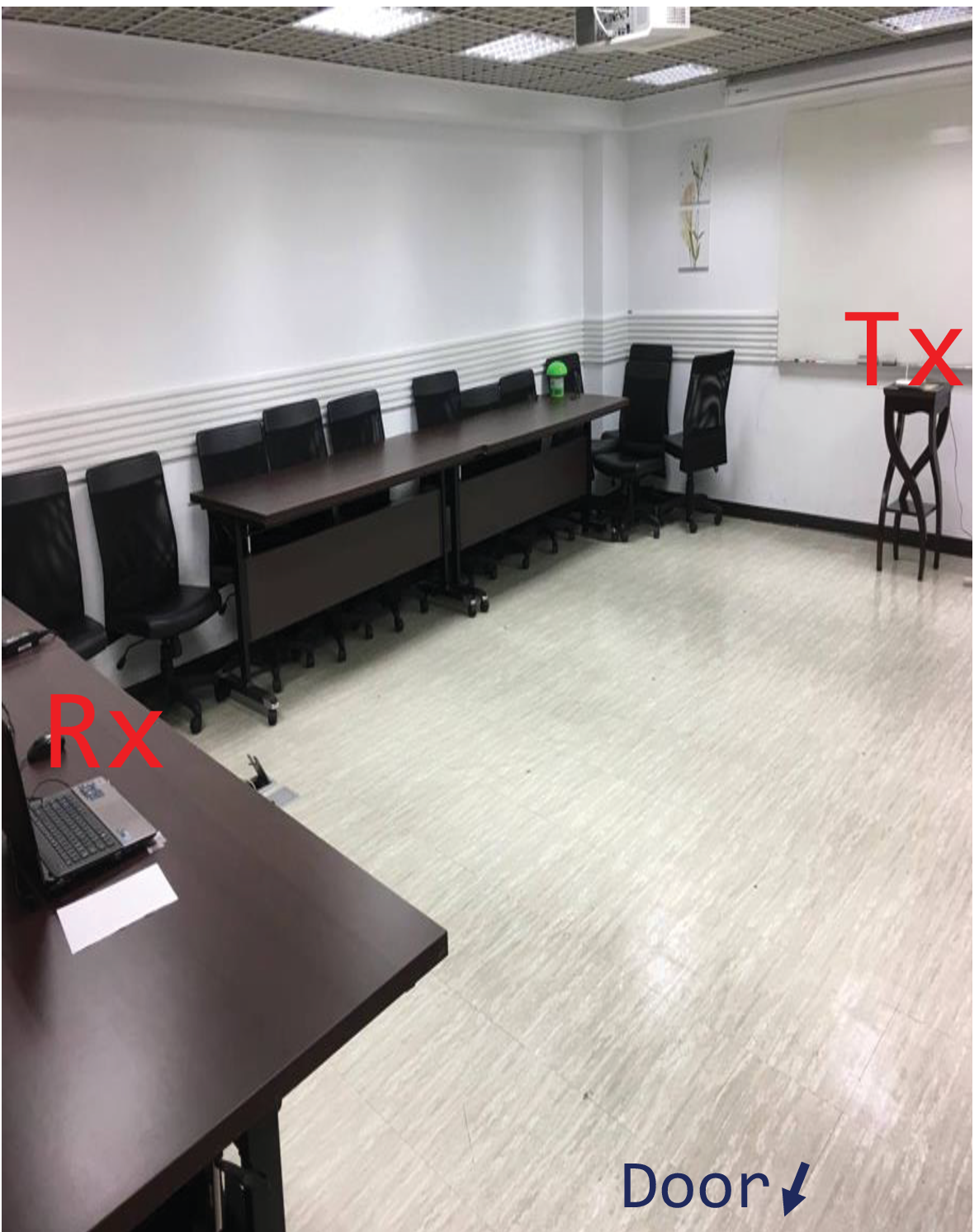}}
\caption{(a) Floor plan and (b) photograph of Scenario A (open-space conference room).}
\label{fig:exp1}
\end{center}
\vspace{-0.25in}
\end{figure}

\begin{figure}[t]
\begin{center}
\subfigure[]{
    \label{exp3 FP}
    \includegraphics[width=0.73\columnwidth]{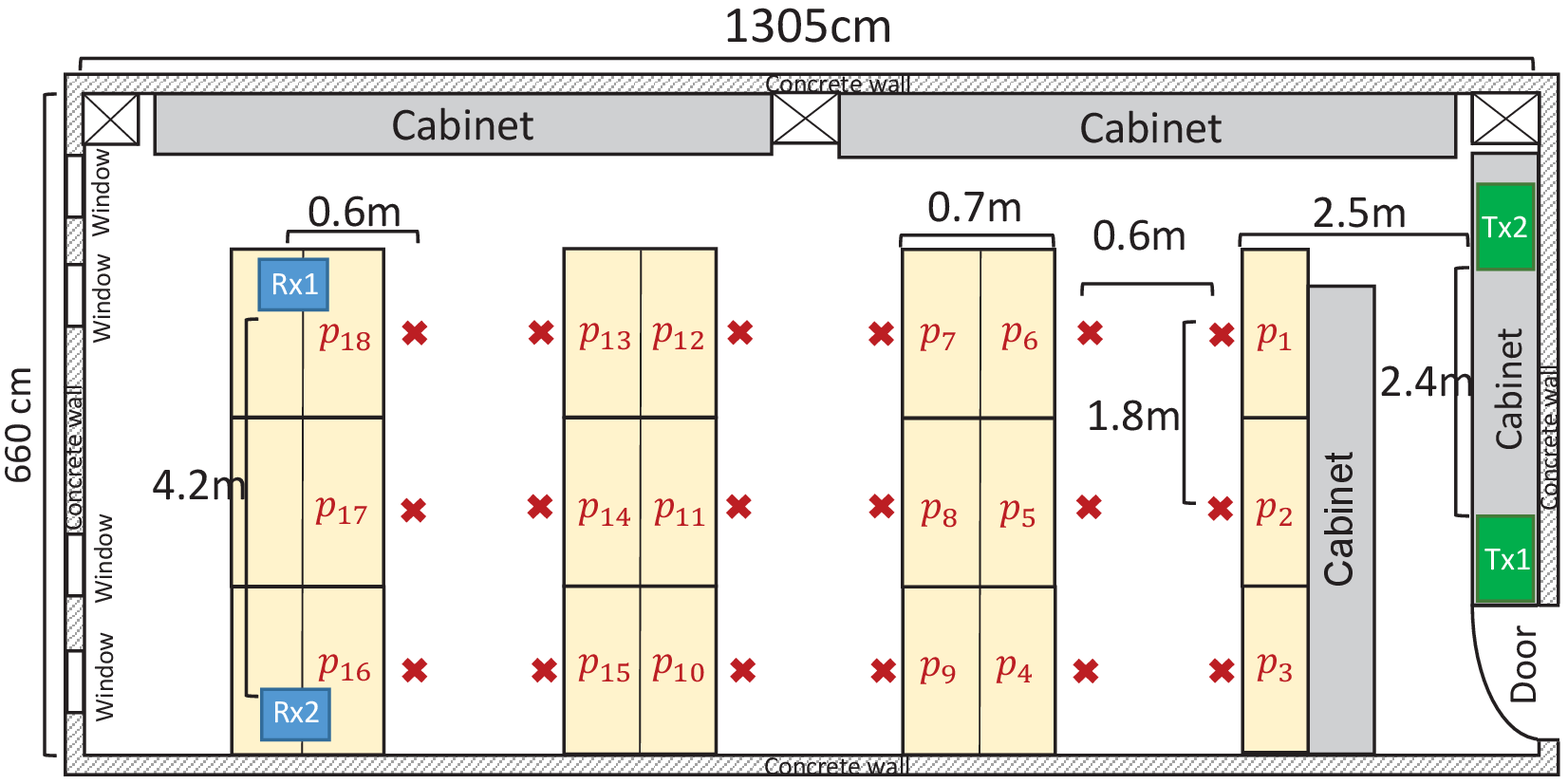}}
\subfigure[]{
    \label{exp3 photo}
    \includegraphics[width=0.708\columnwidth]{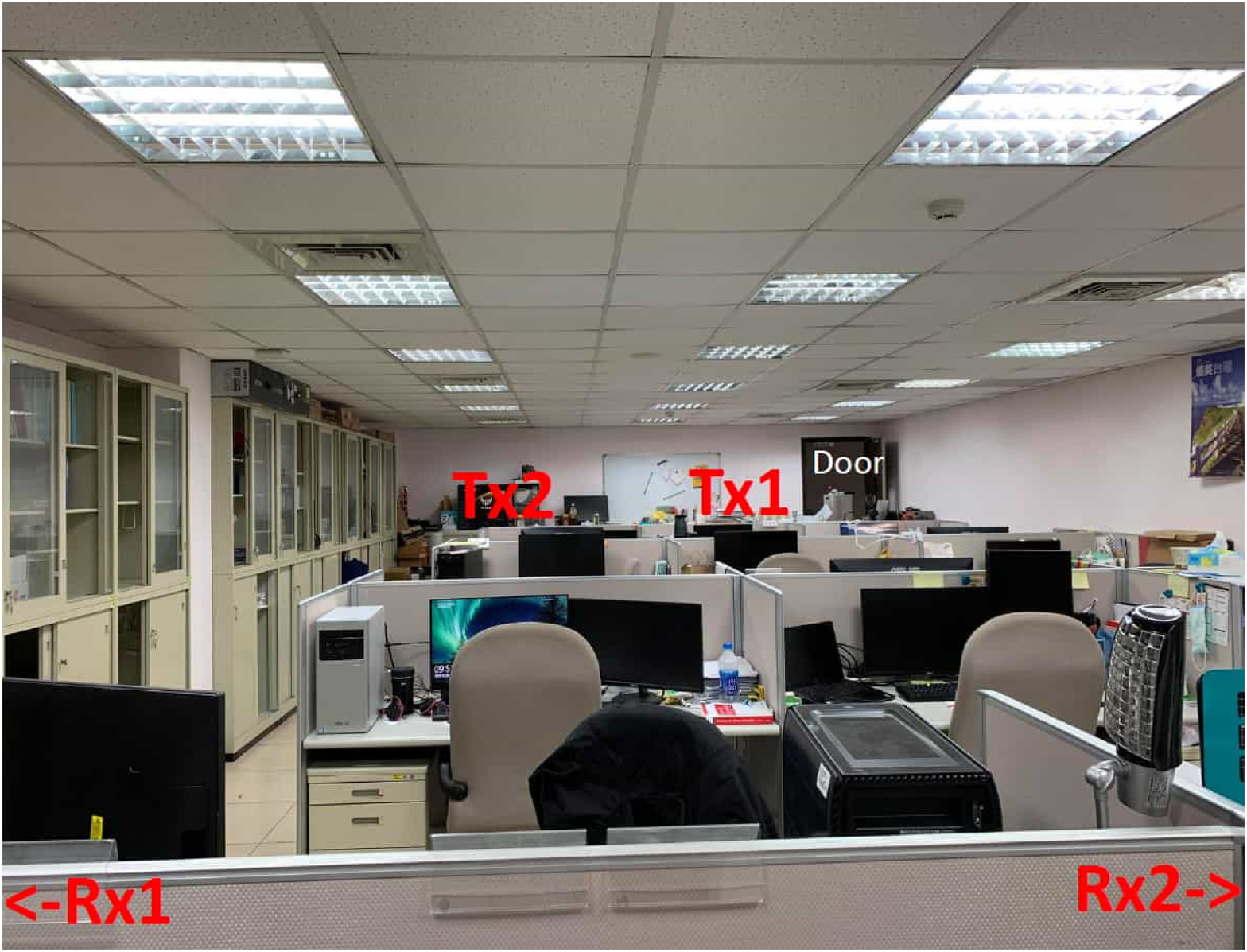}}
\caption{(a) Floor plan and (b) photograph of Scenario B (cubicle office).}
\label{fig:exp3}
\end{center}
\end{figure}

We evaluate the performance of the proposed schemes based on two real-world experiments conducted at the Research Center for Information Technology Innovation, Academia Sinica. Scenario A, as depicted in Fig.~\ref{fig:exp1}, is an open-space conference room with $16$ locations (marked by $p_1, p_2, \ldots, p_{16}$). We collect $600$ CSI samples for each location ($9600$ CSI samples for all locations). Scenario B, as depicted in Fig.~\ref{fig:exp3}, is a cubicle office with $18$ locations (marked by $p_1, p_2, \ldots, p_{18}$). We collect $500$ CSI samples for each location ($9000$ CSI samples for all locations). Scenario A employs one fixed-location transmitter-receiver (Tx-Rx) pair, while Scenario B employs two fixed-location Tx-Rx pairs. Scenario A and Scenario B have completely different layouts, dimensions, numbers of locations, etc., ideal for examining the effectiveness of transfer learning. Both scenarios are device-free.

We examine two cases for transfer learning: 
\begin{itemize}
\item[1)] Scenario A as the source domain and Scenario B as the target domain, and 
\item[2)] Scenario B as the source domain and Scenario A as the target domain. 
\end{itemize}
For Case 1, all the $K_s=600$ samples per location from Scenario A, along with the few $K+1$ samples per location from Scenario B, are used to comprise the training dataset. Here, $M=16$ and $N=18$, and we consider $K=1, 5, 10$. Thus, we compare 18-way 1-shot, 18-way 5-shot, and 18-way 10-shot schemes in the framework of $N$-way $K$-shot learning. The remaining $K_t=500-(K+1)$ samples per location from the target domain which are not part of the training dataset are used for the testing dataset. We randomly select a sample from the testing dataset into the query set to evaluate the performance and average over $6400$ times. For Case 2, we have $M=18$, $N=16$, and $K=1,5,10$. We repeat the same procedure as in Case 1 and compare 16-way 1-shot, 16-way 5-shot, and 16-way 10-shot schemes.

\subsection{Model Settings}

\subsubsection{Feature Extractor} \label{sec:feature extractor setting}

We implement the feature extractor by a CNN with three consecutive 1D convolutional layers for 32 kernels of size 5, strides 2, and zero-padding to halve the dimensions at each layer. Each 1D convolutional layer is followed by batch normalization and ReLU activation. Finally, a fully-connected layer of output size 32 is applied, whose output is our extracted features. In the pretraining (fine-tuning) phase, we concatenate the classifier layer with output size $M$ ($N$) corresponding to the source domain (target domain).

\subsubsection{Graph Convolutional Layers}

The edge update function $f_{\theta}$ comprises three consecutive 2D convolutional layers with output channel sizes being 32, 16, and 1, respectively, and kernels of size 1 and strides 1 for all. The first two 2D convolutional layers are followed by batch normalization and LeakyReLU activation. The node update is performed as in \eqref{eq:node update1} with output size $d'_{\ell-1}=d_{\ell-1}/2$ and $L=2$.

\subsubsection{Classification Layer}

Adam optimizer is used to optimize the model with learning rate $0.01$ and weight decay $10^{-6}$.

\subsection{Performance Comparison and Discussion}

\begin{table}[t]
\caption{Localization Performance (Classification Accuracy in $\%$) in Scenario B (Case 1 for Transfer Learning Schemes)}
\begin{center}
    \begin{tabular}{|c|c|c|c|}
    \hline
         \textbf{}& \textbf{18-way 1-shot}& \textbf{18-way 5-shot} & \textbf{18-way 10-shot} \\ \hline
         CNN & 27.37\% & 53.07\% & 69.61\%  \\ \hline
         GNN & 37.51\% & 66.70\% & 85.25\%  \\ \hline
         Attentive GNN & 44.17\% & 72.05\% & 85.23\%  \\ \hline
         EGNN & 49.47\% & 72.16\% & 87.33\%  \\ \hline
         ChebyNet & 47.19\% & 74.75\% & 85.55\%  \\ \hline
    \end{tabular}
\label{tab:exp1}
\end{center}
\vspace{-0.15in}
\end{table}

\begin{table}[t]
\caption{Localization Performance (Classification Accuracy in $\%$) in Scenario A (Case 2 for Transfer Learning Schemes)}
\begin{center}
    \begin{tabular}{|c|c|c|c|}
    \hline
         \textbf{}& \textbf{16-way 1-shot}& \textbf{16-way 5-shot} & \textbf{16-way 10-shot} \\ \hline
         CNN & 29.00\% & 51.47\% & 72.59\%  \\ \hline
         GNN & 63.14\% & 72.11\% & 87.78\%  \\ \hline
         Attentive GNN & 70.69\% & 84.16\% & 89.45\%  \\ \hline
         EGNN & 66.75\% & 87.23\% & 90.44\%  \\ \hline
         ChebyNet & 73.69\% & 85.42\% & 88.52\%  \\ \hline
    \end{tabular}
\label{tab:exp3}
\end{center}
\end{table}

Table~\ref{tab:exp1} summarizes the localization performance measured by the classification accuracy in Scenario B. The CNN scheme shown for comparison is the same as the feature extractor described in Sec.~\ref{sec:feature extractor setting} plus an output classification layer of size $N$. CNN does not employ transfer learning, but we slightly abuse the notation and use $N$-way $K$-shot to refer to the CNN scheme trained with $K+1$ samples per location from the target domain without any samples from the source domain. Table~\ref{tab:exp3} summarizes the localization performance in Scenario A.

As can be seen from Table~\ref{tab:exp1} and Table~\ref{tab:exp3}, first, GNN-based few-shot transfer learning models significantly outperform CNN. This illustrates the effectiveness of transfer learning by exploiting the existing samples from a different environment (i.e., the source domain). The performance boost is because the models can learn abstraction from the source domain which is useful for the target domain. Second, the performance of all schemes increases as $K$ increases. Increasing $K$ beyond $10$ yields diminishing returns and is uncommon in the framework of few-shot learning. Third, the performance boost due to transfer learning is higher in Case 2 than in Case 1, especially in the very few shot (1-shot and 5-shot) regimes. This may be explained by the fact that the layout is simpler and the data samples are cleaner in Scenario A than in Scenario B. Transferring the abstraction learned from a more complex environment (Scenario B) to a simpler environment (Scenario A) as in Case 2 allows the model to adapt to the new environment more quickly and results in a better performance. Fourth, the three proposed modifications to the original GNN model all improve the GNN by various, yet comparable, degrees. The improvement is particularly noticeable in the very few shot (1-shot and 5-shot) regimes. The improvement is relatively small in the 10-shot regime because as $K$ increases, the size of the fully-connected graph increases dramatically, and the issue of over-smoothing becomes more moderate since the information exchange is more complicated and the node features are no longer monotonic.

To get an idea of the performance shown in Table~\ref{tab:exp1} and Table~\ref{tab:exp3} in absolute terms, we train an independent CNN model with a large amount of data ($400$ samples per location) for both scenarios. The resulted performance is $95.83\%$ and $85.91\%$ in Scenario B and Scenario A, respectively. As can be seen, the proposed schemes achieve close performance in Scenario B and even better performance in Scenario A using significantly reduced numbers of samples ($40$ times fewer) in the target domain. This confirms the merit of the proposed method in fingerprint-based indoor localization, where data previously collected in other environments are reused and the model learned therein is transferred to a new environment to ease the data collection/labeling effort in the new environment.

\section{Conclusion} \label{sec:conclusion}

In this paper, we have proposed a GNN-based few-shot transfer learning system for device-free fingerprinting indoor localization. The proposed system requires only very few CSI samples from the target domain. The proposed system presents a general method to reduce the data-labeling effort by reusing the data from any other, possibly very different, environments, as verified in our experiments with two heterogeneous scenarios. A mathematical description of the proposed system was presented and the comparison of different modifications was discussed. The proposed system provides an effective solution to the common challenges faced by fingerprint-based approaches, i.e., fingerprint database collection, maintenance/update, and transfer.

\bibliographystyle{IEEEtran}
\bibliography{IEEEabrv,ref}

\end{document}